\documentstyle[aps,prl,epsfig,twocolumn]{revtex}
\renewcommand{\vec}[1]{\relax\ifmmode\mathchoice
{\mbox{\boldmath$\relax\displaystyle#1$}}
{\mbox{\boldmath$\relax\textstyle#1$}}
{\mbox{\boldmath$\relax\scriptstyle#1$}}
{\mbox{\boldmath$\relax\scriptscriptstyle#1$}}\else
\hbox{\boldmath$\relax\textstyle#1$}\fi}
\begin{document}
\pagestyle{myheadings}
\markboth{Helbing/Farkas/Vicsek: Freezing by Heating}
{Helbing/Farkas/Vicsek: Freezing by Heating}
\tighten
\onecolumn
\twocolumn[\hsize\textwidth\columnwidth\hsize\csname @twocolumnfalse\endcsname

\title{Freezing by Heating in A Driven Mesoscopic System}

\author{Dirk Helbing$^{*,+}$, Ill\'{e}s J. Farkas$^+$, and Tam\'{a}s Vicsek$^+$}
\address{$^*$ II. Institute of Theoretical Physics, University of Stuttgart,
Pfaffenwaldring 57/III, 70550 Stuttgart, Germany\\
$^+$ Department of Biological Physics, E\"otv\"os University, Budapest,
P\'azm\'any P\'eter S\'et\'any 1A, H-1117 Hungary\\
{\tt helbing@theo2.physik.uni-stuttgart.de; fij@elte.hu;
vicsek@angel.elte.hu}} 

\maketitle

\begin{abstract}

We investigate a simple model corresponding to particles driven
in opposite directions and
interacting via a repulsive potential.  The particles
move off-lattice on a periodic strip and are subject to random forces as well.
We show that this model---which can be considered as a continuum version
of some driven diffusive systems---exhibits a paradoxial, new
kind of transition called here ``freezing by heating''. One
interesting feature of this transition is that a {\it crystallized state}
with a higher total energy is obtained from a fluid state
{\it by increasing the amount of fluctuations}.

\end{abstract}
\pacs{PACS numbers: 05.70.Fh, 64.60.Cn, 05.70.Ln}
]
Most of the phenomena in our natural environment occur under far-from
equilibrium conditions resulting in a rich behavior both in time and
space.  An important class of such processes takes place in
so-called driven systems, which have attracted considerable interest
recently. In many of these systems, particles are driven either by an external
field (force) \cite{gran,order,fluid},
or they are self-propelled
\cite{Vics,mot}, 
and their collective behavior manifests itself
in new kinds of 
transitions, including noise
induced ordering \cite{order1,order} 
or an ordering in
a continuous 2d velocity space\cite{Vics}.
\par
Phase transitions are also common in equilibrium systems, and the related
analogies have represented an important contribution to the
understanding of non-equilibrium processes.  In most cases,
lattice models have been considered to demonstrate
non-equilibrium transitions.  For example, jamming transitions have been
seen in discretized traffic models \cite{traf} and
driven lattice gases \cite{driven}.
However, off-lattice (i.e., continuum) symmetry is known to bring in
qualitatively new behaviour; in particular, this is definitely so in 2d,
see, e.g., the XY model\cite{KT} versus the Ising model (in
equilibrium). A
continuum model may lead to new effects due to the fact that the notions
of order and disorder have extra facets in this case, and it can
describe compressible systems in a more delicate way.
\par
In this paper we will consider a simple continuum model
exhibiting a paradoxial, new kind of transition that we call
``freezing by heating''
and being closely related to
situations relevant from the practical point of view.  The model
consists of particles driven in opposite directions and interacting
through a simple repulsive potential.  The particles move off-lattice on
a periodic strip (in a two-dimensional tube) 
and are subject to random forces as well.  The most
interesting feature of the transition we find for this system is, that a
{\it crystallized state} with a higher total energy is
achieved from a fluid state over a transient disordered state
{\it by increasing the amount of fluctuations}.
\par
In addition to the interest in the properties of driven systems on its own,
there are several further motivations to study such a model.
A system of
light (rising) and heavy (sinking) particles in a vertical column of fluid,
pedestrians moving in a passage, or a system
of oppositely charged colloidal particles in an electric field
represent potential applications of our model.  In fact, the
system we study is a generalization to the continuum case of a
two-species driven lattice gas model proposed recently \cite{SHZ}, with a
number of relevant modifications arising from the adaptation to the
off-lattice case.  In a wider context, these models can be considered as
simplified paradigms of systems consisting of entities with opposing
interests (drives).  In the present work, we consider the behavior of a
limited number of particles in a confined geometry, and our results are
primarily valid for this ``mesoscopic'' situation.  In the quickly growing
literature on mesoscopic systems there are many examples of the
potential practical relevance of phenomena occurring in various
models for finite sizes \cite{meso}.

We denote the location of
particle $i$ at time $t$ by $\vec{x}_i(t)$ and its
velocity $d\vec{x}_i(t)/dt$ by $\vec{v}_i(t)$. Furthermore, we assume
the acceleration equation
\begin{eqnarray}
m \frac{d\vec{v}_i(t)}{dt}
&=& m \frac{v_0 \vec{e}_i -\vec{v}_i(t)}{\tau} + \vec{\xi}_i(t)
 \nonumber \\
&+& \sum_{j (\ne i)} \vec{f}_{\!ij}
 \Big(\vec{x}_i(t),\vec{x}_j(t)\Big)
+ \vec{f}_{\rm b}\Big(\vec{x}_i(t)\Big) \, .
\label{motion}
\end{eqnarray}
$m$ is the mass of the particle, $v_0$ the velocity with which it tends to
move in the absence of interactions, $\tau$ the corresponding relaxation
time, and $\vec{e}_i \in
\{(1,0),(-1,0)\}$ the direction into which particle $i$ is
driven. $\gamma = m/\tau$ may be interpreted as a friction coefficient.
$\vec{f}_{\!ij}$ represents the repulsive interactions
between particles $i$ and $j$, $\vec{f}_{\rm b}$ the interactions
with the boundaries, and $\vec{\xi}_i$ the fluctuations
of the individual velocities. For the
interactions between the particles, we have chosen the simple function
\begin{eqnarray}
 \vec{f}_{\!ij} (\vec{x}_i,\vec{x}_j)
 &=& - \vec{\nabla} A (d_{ij}-D)^{-B} \, ,
\label{inter}
\end{eqnarray}
depending on the parameters $A$ and $B$, and
the distance $d_{ij}(t) = \|\vec{x}_i(t) - \vec{x}_j(t)\| > D$
only. Thus,
$\vec{f}_{\!ij}$ describes the effect of a
soft repulsive potential of particle $j$ with a hard
core of diameter $D$, reflecting the space occupied by the
particle.
Our choice of these details of the model corresponds to a
motion of
finite sized particles tending to avoid collisions and maintaining, if
possible, a given velocity $v_0$. 
In addition, the interactions with the boundaries were assumed to be
\begin{eqnarray}
 \vec{f}_{\rm b}(\vec{x}_i) &=& - \vec{\nabla} A (d_{i \perp} - D/2)^{-B}
 \, ,
\label{wall}
\end{eqnarray}
where $d_{i\perp}$ denotes the shortest distance to the closest
wall.
\par
\begin{figure}
\begin{center}
\begin{picture}(8,6.4)
\put(0.3,4.4){\epsfig{width=8\unitlength, angle=0,
      file=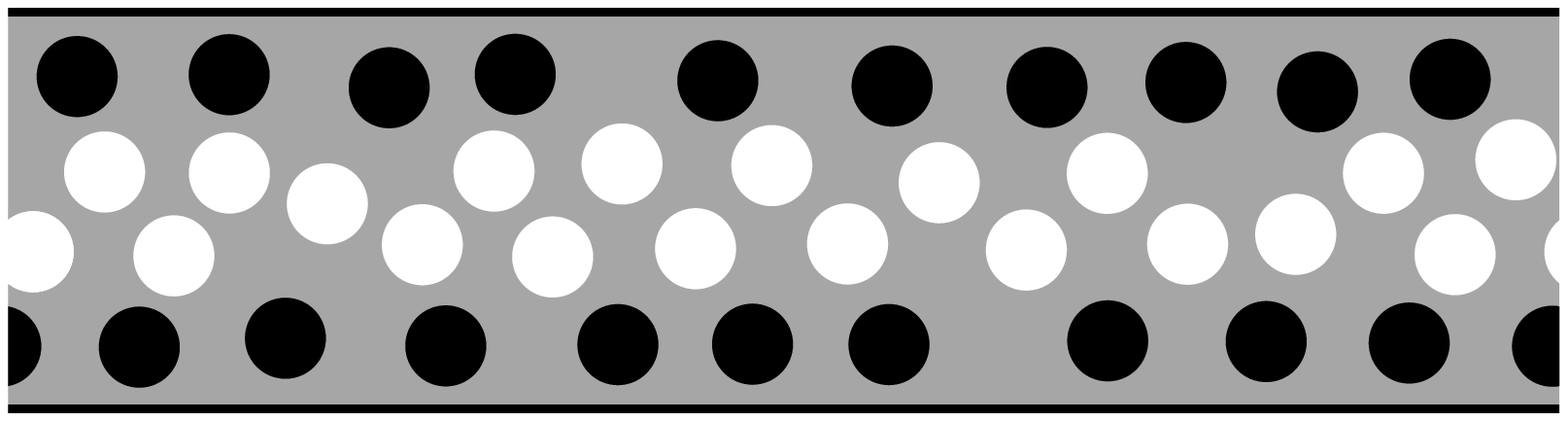}} 
\put(0.3,2.2){\epsfig{width=8\unitlength, angle=0,
      file=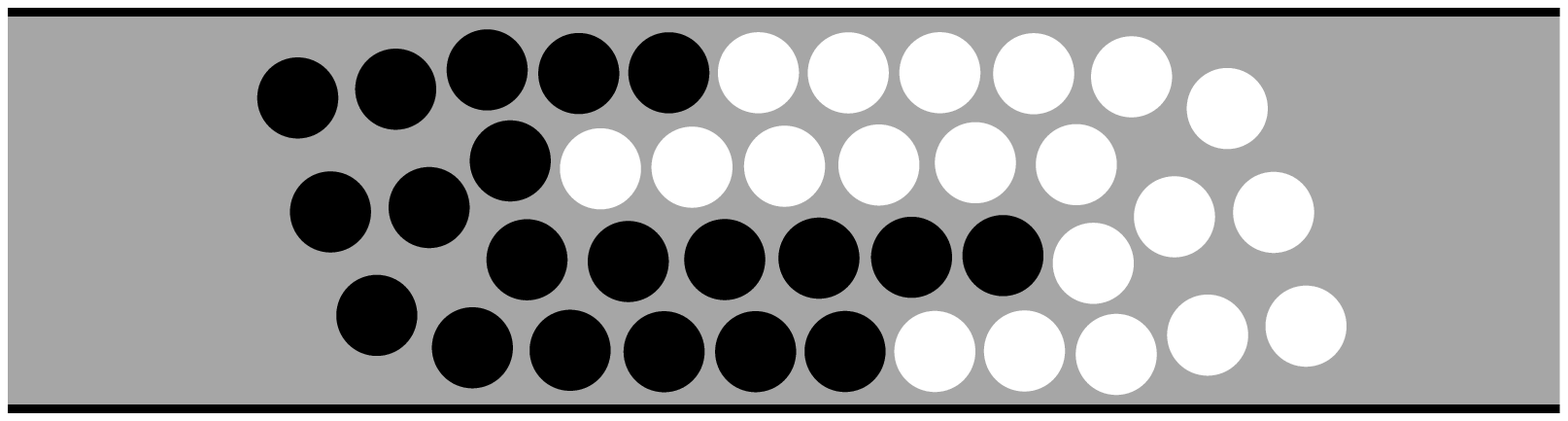}}
\put(0.3,0){\epsfig{width=8\unitlength, angle=0,
      file=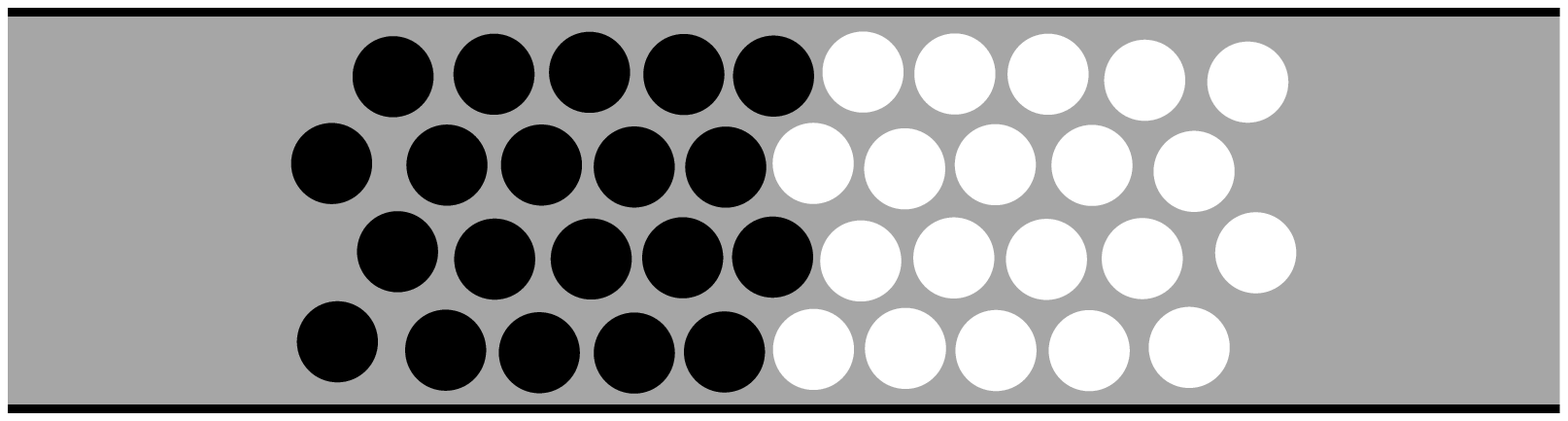}}
\put(-0.3,6.2){(a)}
\put(-0.3,4.0){(b)}
\put(-0.3,1.8){(c)}
\end{picture}
\vspace*{0.5\unitlength}
\end{center}
\caption[]{Simulation of 20 particles moving
from left to right (black)
which interact with 20 particles moving 
from right to left (white) on a periodic
strip of length $L_x=20$ and width $L_y=5$ at different noise
intensities. The model parameters are $m=1$, $D=1$, $v_0 = 1$,
$A = 0.2$, $B = 2$, and $\tau = 0.2$. 
(a) Lanes of uniform directions of motion forming at small noise intensity
($\theta = 1$). 
(b) Snapshot of an intermediate jammed state with a rough interface, which
is about to form ``channels''. 
(c) Final crystallized state resulting for large noise intensity
($\theta = 1000$).
\label{fig1}}
\end{figure}

In contrast to previous studies of similar models \cite{ped}, we
will now investigate the
decisive role of the fluctuations $\vec{\xi}_i(t)$, which
have been assumed to be uncorrelated and distributed according to a
truncated normal distribution with vanishing mean value and finite variance
$\theta$ \cite{note}.
We started our simulations with N particles
randomly distributed on a strip without allowing overlaps. For
half of the particles a driving into the $(-1,0)$ direction, and for the other
half a driving into the $(1,0)$ direction was assigned.
Numerical integration of equation (\ref{motion}), using
periodic boundary conditions, has produced
the following results:  For small noise amplitudes 
$\theta$ and sufficiently small particle
densities, our simulations lead, depending on the strip width and
the initial condition, to the formation of two or more
coherently moving linear structures (just as if the
particles moved along traffic lanes)
(cf.  Fig.~\ref{fig1}a).  For relatively large
$N$ (if the available area is too small to allow freely moving
lanes), jamming occurs. For a small intermediate density region,
we find lane formation or jamming, depending on the respective
initial condition.
\par
At small noise amplitudes,
the mechanism of lane formation, which produces a ``fluid'' state,
is very dominant and robust in our model. 
This can be understood as follows: Particles
moving against the stream or in areas of mixed directions of motion
will have frequent and strong interactions, because of high relative
velocities. In each interaction, the encountering
particles move a little aside to pass each other.  This
sidewards movement tends to separate oppositely moving particles.
Nevertheless, jamming may sometimes occur, but in most cases it also
supports lane formation (see next paragraph). Particles moving in uniform lanes
have very rare and weak interactions.  Hence, the tendency to break up
existing lanes is negligible, when the fluctuations are small.
Furthermore, the most stable configuration corresponds to a state with a
minimal interaction rate and is related with a maximum efficiency
of motion \cite{sci}.
\par
Whereas {\em spontaneous}
lane formation was also observed in previous studies of related
models with {\em deterministic} dynamics only \cite{ped}, in the
present, more realistic model we have discovered a surprising phenomenon
when we increased the noise amplitude. 
If the fluctuations and
the particle number are large
enough, the particles crystallize into a hexagonal lattice. This is a
consequence of several subsequent steps: First, the fluctuations
are able to prevent lane formation or even to destroy
previously existing lanes. This is so, because sufficently
strong diffusion can prevent structure formation.
Second, some of the oppositely moving particles
block each other locally from time to time.
Third, this gives rise to jamming since, meanwhile,
additional particles arrive at the boundaries of the
blocked area. Fourth, if the jam exists long enough,
both of its ends expand over the full width of the strip and
develop ``flat'' boundaries perpendicular to $\vec{e}_i$, in order to
reach a balance of forces. For the same reason, the
particles tend to arrange in a hexagonal lattice structure, very much like in a
crystal. Fifth, the crystal is only stationary, 
if also the {\em interface} between
the oppositely moving particles is, by chance, flat enough (cf.  
Fig.~\ref{fig1}c).
In most cases, however, the interface is rough (cf.  Fig.~\ref{fig1}b),
i.e. in some of the horizontal layers, a
{\em majority} of particles is pushing in {\em one} direction. As a
consequence, the most advanced part(s) of the interface eventually break(s) 
through, which requires a continuous model, where the distance kept among the
particles is flexible enough. In this way, particles with uniform
directions of motion form ``channels'', which tend
to produce lanes at sufficiently small densities and noise intensities,
otherwise the particles jam again and again (as described  above),
until they end up in a {\em stationary} crystal.
\par
Due to the above described mechanism the
crystallized state is {\em metastable}, i.e., sensitive
to structural perturbations (like the interchange of a few particles
in our case). The crystallized state can also be 
destroyed by {\em ongoing} fluctuations with extreme noise amplitudes
giving rise to a third, disordered (``gaseous'')
state with randomly distributed particles. 
Thus, with increasing ``temperature'' $\theta$, we have the untypical
sequence of transitions {\em fluid $\rightarrow$ solid $\rightarrow$
gaseous}.
\par
\begin{figure}
\begin{center}
\begin{picture}(8,5.5)
\put(-0.3,6.2){\epsfig{width=6.2\unitlength, angle=-90,
      file=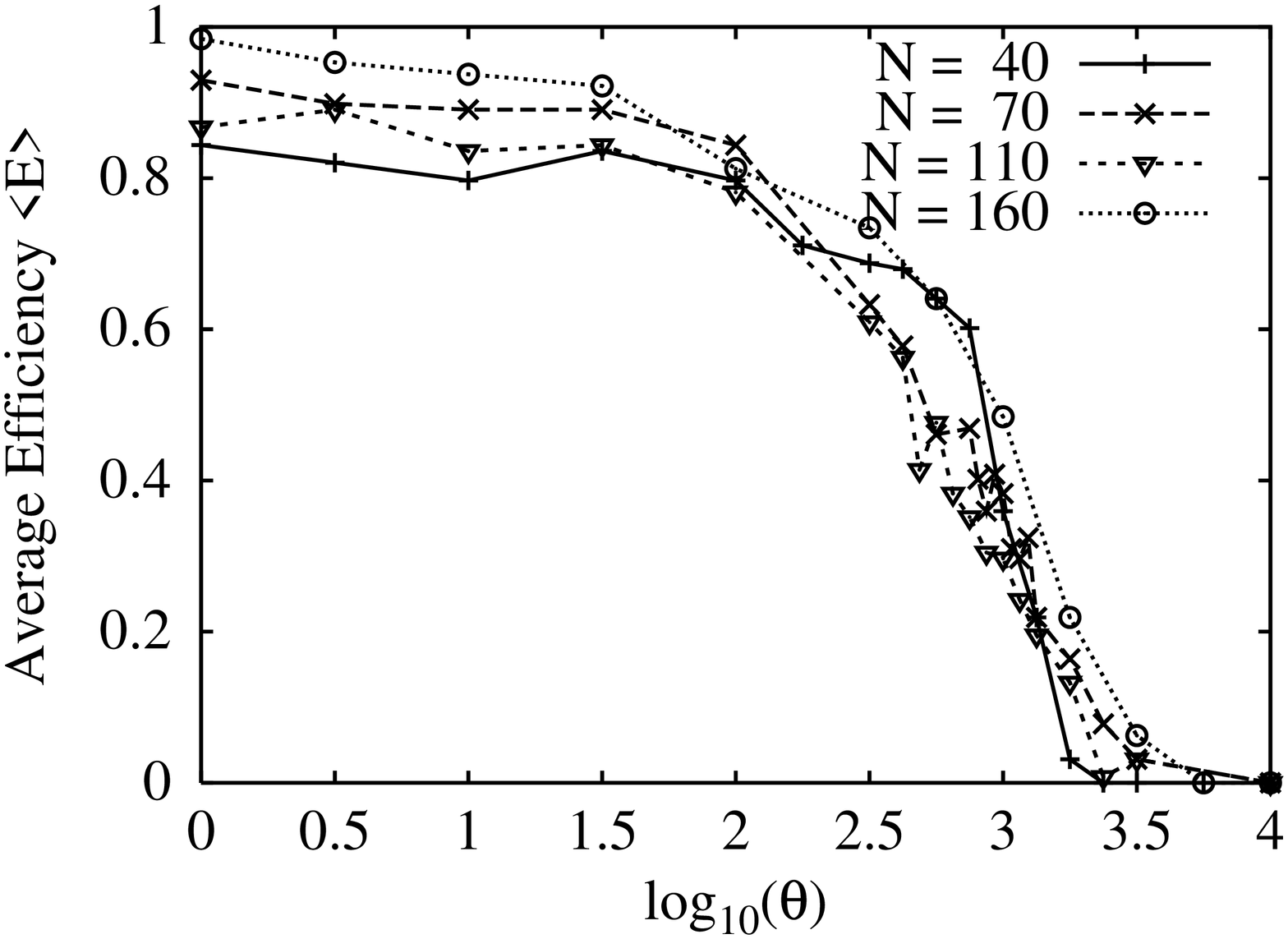}}
\put(1.5,4.0){\epsfig{width=2.7\unitlength, angle=-90,
      file=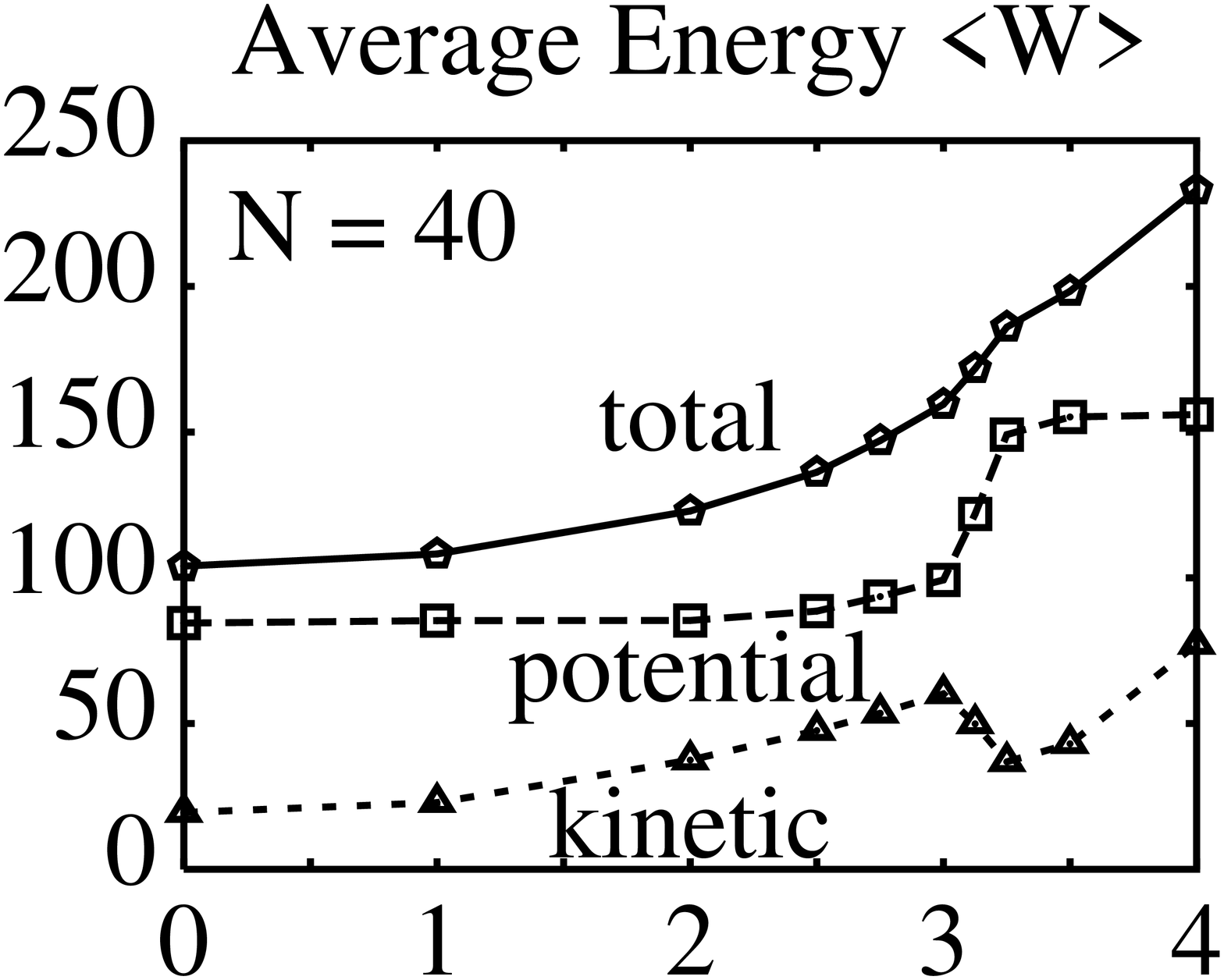}}
\end{picture}
\end{center}
\caption[]{Efficiency $\langle E \rangle$ of the system as a function
of the noise intensity $\theta$ for various system sizes with
a fixed aspect ratio of $L_x/L_y=4:1$ and the parameter values displayed
in Figure~\ref{fig1}. Shown are averages over 128 (for $N=160$: 64)
simulation runs with different random seeds. The inset shows the
average potential, kinetic, and total energy in
the resulting system states.
While the potential energy  (--~--) 
increases with the noise amplitude 
due to jamming,
the kinetic energy  (-~-~-) is composed of a contribution proportional to
$\theta$ and a decreasing contribution $N m (v_0 \langle E
\rangle)^2/2$. The total energy  (---) turns out to be an increasing function 
of $\theta$.
\label{fig2}}
\end{figure}

Interestingly, for a range of moderate densities we
find a ``fluid'' state with lanes at small noise amplitudes 
most of the time, but a crystallized (``frozen'') state,
if the noise amplitude 
(``temperature'') is large. 
We call this transition ``freezing by
heating''. Starting with {\em random} initial conditions,
the transition is rather smooth (cf.  Fig.~\ref{fig2}). 
This is partially so because the system can also become 
frozen at a relatively {\em low} noise amplitude, 
if the {\em disorder} in the
initial state (in the sense of the deviation from a freely moving lane state) 
is large enough (which has an effect similar to additional fluctuations). 
The transition becomes sharper, if we always start
with a two-lane state but with different random seeds.
In any case, the transition is {\em hysteretic}, since the
noise-induced ``frozen'' state remains, when the noise amplitude 
is  reduced, again. 
\par
To characterize the state of the system, we calculated various
quantities. The expression
\begin{equation}
 E = \lim_{T\to \infty} \frac{1}{T} \int\limits_0^T \! dt \;
 \frac{1}{N} \sum_{i=1}^N \frac{\vec{v}_i(t)\cdot \vec{e}_i}{v_0} \, ,
\end{equation}
for which we expect the relation $0 \le E \le 1$, is
a measure for the ``efficiency'' of motion, i.e., $E v_0$ is the
average speed at which the
particles are able to move in their respective ``target direction''
$\vec{e}_i$. $E\approx 1$ corresponds to lanes, $E=0$ to a
crystallized state.
Representative simulation results for the ensemble average
$\langle E \rangle$ as a function of
the noise intensity $\theta$ are displayed in Figure~\ref{fig2}.
\par
We observed the following parameter dependencies: Crystallization is
more pronounced for large $\tau$ and large strip lengths $L_x$,
while small $\tau$ and large strip widths $L_y$ are in favour of
lane formation (but ``freezing by heating'' still exists in the
overdamped limit $\tau \rightarrow 0$). The number
of particles required for crystallization does not depend on the
length $L_x$ (if it is considerably larger than $L_y$),
while it is roughly proportional to the width $L_y$ (for large enough
$L_y$). Given a fixed aspect ratio $L_x/L_y$, for the system sizes
that we could numerically handle
there was no clear tendency whether the transition becomes sharper 
or smoother with increasing system size $N \propto L_x L_y$ 
(see Fig.~\ref{fig2}).
\par
Another interesting quantity is the sum of
the potential and kinetic energies associated with a given state:
\begin{equation}
W = \lim_{T\to \infty} \frac{1}{T} \int\limits_0^T \! dt \; \Bigg[
 \sum_i \frac{m}{2} \, \vec{v}_i^2
 + \frac{1}{2} \sum_{i\ne j} A (d_{ij}-D)^{-B} \Bigg] \, .
\label{inten}
\end{equation}
The paradox here is that the above mentioned crystallized state is
usually more unstable than the fluid state,
in the sense that the total energy (\ref{inten})
of the system is higher in the crystallized state than
in the ``fluid'' one (see inset of Fig.~\ref{fig2}). Note that both,
the ``solid'' (crystallized) state and the ``fluid'' state (i.e. lanes) are
destabilized, if the friction term $-\vec{v}_i/\tau$ is dropped, even
in the case $\theta = 0$ (see Ref.~\cite{extra} for a related inverse
phenomenon). That is, without friction and due to the
permanent driving, the undamped repulsive
interactions eventually become destructive to any ordered state, 
which gives rise to a ``gaseous'' state.
Therefore, we point out that
the energetically less favourable crystallized state is maintained
by the propulsion term $m (v_0\vec{e}_i - \vec{v}_i)/\tau$ 
in Eq. (\ref{motion}) which, by the way, is also relevant for lane formation.
Note that the absolute value of this term becomes largest for
$\vec{v}_i = \vec{0}$ (i.e. blocking), while it is small for the
``fluid'' state with $\vec{v}_i \approx v_0 \vec{e}_i$.
\par
We consider the transition to a stationary state with a higher total
energy by increasing the noise intensity to be a signature of a novel class of
behavior in certain non-equilibrium systems,
which may have interesting applications.
However, here we have demonstrated ``freezing by heating'' only for limited
sizes and a specific geometry. Nevertheless, we point out that ``freezing by
heating'' does not require walls, but occurs for periodic boundary
conditions in $y$-direction as well. Also, we do not need periodic boundary
conditions in $x$-direction. Any sufficiently long simulation area
will produce both the organized fluid state (i.e. lanes) and the
crystallized state, if only the system is continuously entered 
by particles at the left-hand and right-hand boundaries.
\par
Why is freezing by heating new?  Some glasses may crystallize when
slowly heated.  However, this is a well understood phenomenon: the
amorphous state is metastable for those temperatures, and
crystallization means an approach to the more stable state with smaller
total energy.
In general, one can distinguish three cases when fluctuations (i.e.
temperature or external perturbations) are increased: (i) Total energy
increases and order is destroyed (e.g., melting). (ii) Total energy
decreases, ordering takes place, and the system goes from a disordered
metastable to an ordered stable state (e.g., in metallic glasses and some
granular systems). (iii)
Total energy increases and ordering takes place, while the system goes
from a partially ordered stable to a highly ordered metastable state, which
corresponds to the new situation presented here.
\par
In our case, crystallization is
achieved by spontaneously driving the system with the help of noise
uphill towards higher total energy.  The system would like to
maximize its efficiency \cite{sci}, but instead it ends up with minimal
efficiency due to noise-induced crystallization.  The role of
``temperature'' or noise here is to destroy the energetically more
favourable fluid state, which inevitably leads to jamming and finally to
crystall-like lattices. The corresponding transition seems to be
related to the off-lattice nature of our model and is different from
those reported for driven diffusive systems on a lattice
\cite{driven,SHZ}.  It should be
noted that the transition we find is not sharp, which is a
consequence partly of the mesoscopic nature of the phenomenon and
partly of the disorder in the initial state.
\par
We would like to point out that ``freezing by heating'' is likely to be
relevant to situations involving pedestrians under extreme conditions
(panics).  Imagine a very smoky situation, caused by a fire, in which
people do not know which is the right way to escape.  When panicking,
people will just try to get ahead, with a reduced tendency to follow a
certain direction.  Thus, fluctuations will be very large, which can
lead to fatal blockings.
\par
Our results demonstrate that in driven mesoscopic systems phenomena
qualitatively different from those occurring in thermodynamical systems
can be observed.
While most non-equilibrium transitions have analogies to equilibrium
ones\cite{phasetrans}, the
noise-induced ordering observed in the effect of
``freezing by heating'' is just opposite to
the transitions occurring in equilibrium systems.
This suggests that future studies along the lines
of the present approach are likely to lead to further unexpected
findings.

{\em Acknowledgments:}
D.H. wants to thank the DFG for financial support
by a Heisenberg scholarship and D. Mukamel for helpful suggestions.
This work was supported by OTKA F019299 and FKFP 0203/1997.

\unitlength10mm 
\unitlength10mm 

\begin{references}
\bibitem{note} We limited the fluctuations to a maximum of
$\|\vec{\xi}_i\| = 3 \sqrt{\theta}$. 
Truncation at larger fluctuation levels gave
practically the same results, but at the cost of considerably
decreased computational efficiency.
\end{references}
\end{document}